\documentstyle[aps,preprint,pra]{revtex}
\tighten
\input{boxedeps.tex}
\SetRokickiEPSFSpecial
\HideDisplacementBoxes
\draft
\begin{document}
\title{Bound-Free Electron-Positron Pair Production in Relativistic Heavy Ion
 Collisions }
\author{Helmar Meier, Zlatko Halabuka, Kai Hencken, and Dirk Trautmann}
\address{
Institut f\"ur theoretische Physik der Universit\"at Basel,
Klingelbergstrasse 82, 4056 Basel, Switzerland}
\author{Gerhard Baur}
\address{Institut f\"ur Kernphysik (Theorie), Forschungszentrum J\"ulich, 
52425 J\"ulich, Germany}
\date{\today}
\maketitle
\begin{abstract}
We study the bound-free electron-positron pair production in
relativistic heavy ion collisions to different bound states in a full
Plane Wave Born Approximation (PWBA) calculation. Exact Dirac wave
functions are used for both the bound electron and the free positron
in the final states. Results for the Relativistic Heavy Ion Collider
RHIC as well as the forthcoming Large Hadron Collider (LHC) are
given. This process is one of the dominant beam loss processes and can
become critical for the operation with heavy ions. A simple
parameterization is given as well.  We compare our results with
calculations of other groups.
\end{abstract}

\section{Introduction}

Bound free pair production is one of the new types of processes that
occur in relativistic collisions of atoms and ions. It is the
production of an electron-positron pair with the electron not produced
as a free state but as a bound state of one of the ions:
\begin{equation}
Z_P+Z_T\rightarrow Z_P+(Z_T+e^-)+e^+.
\end{equation}

This process changes the charge state of the ion. Due to the change in
the charge-to-mass ratio, such an ion will be lost from the
circulating beam and the luminosity will be seriously affected.
Together with the electromagnetic dissociation this process is the
dominant beam loss process at RHIC and the Large Hadron Collider (LHC)
at CERN when using heavy ions \cite{RHIC89,ALICE94,ALICE95}.  In
addition to the beam loss itself, it was recently discussed in
\cite{Klein00} that the capture process can lead to localized beampipe
heating. This could cause magnet quenches if the local cooling is
inadequate. Therefore it is very important to know this cross section
with high accuracy and reliability.

We want to mention also that the corresponding process using
antiprotons as the ``target'' has recently found an application in the
production and detection of relativistic antihydrogen, see
\cite{BaurO96,Blanford98}.  In a collision of the antiprotons with a
target again electron positron pairs are produced. The positron can be
produced in a bound state of the antiproton forming an antihydrogen
atom.  Using the equivalent photon method, this cross section was
calculated in \cite{MungerBS94}.  In the equivalent photon
approximation, the cross section factorizes into a photo-induced cross
section and an equivalent photon spectrum.  This spectrum depends on a
cut-off parameter which should be chosen appropriately.  Using the
PWBA expression for the cross section one can avoid this ambiguity and
see also how this parameter should be chosen. This was done in
\cite{MeierHHT98} and \cite{BertulaniB98}.  In these papers,
differential as well as total cross sections are given as a function
of the collision energy. Whereas in the latter reference approximate
lepton wave functions appropriate for low values of $Z_T$, the charge
of the antiproton, have been used, exact Dirac wave functions, which
are valid also for higher values of $Z_T$, have been used in
\cite{MeierHHT98}.

It is the purpose of this paper to calculate the cross-sections for
the energy region of the colliders RHIC and LHC exactly within the
plane wave Born approximation (PWBA) or (which is equivalent) the
semiclassical straight line approximation (SCA), including also higher
shells.  Previously, such kinds of calculations were done using the
Weizs\"acker-Williams method \cite{AsteHT94,AggerS97}. Higher order
effects (in the interaction with the projectile) have been considered
by a number of groups, see, for example,
\cite{BaltzRW91,BaltzRW93,Becker87}. Recently exact calculations have
been done in the high energy limit by Baltz \cite{Baltz97}. He finds
that that the contributions from higher orders are rather small (of
the 1\% level) and even tend to decrease the cross section.  This is
in contrast to results found at much smaller energies with Lorentz
factors in the target rest frame $\gamma_T<3$
\cite{MombergerGS91,RumrichSG93}, but see also
\cite{BaltzRW94,MombergerBS95}.

In Sec.~\ref{sec_pwba} we extend the formalism presented in
\cite{MeierHHT98} to the relativistic heavy ion case, including now
also formulae for the higher atomic states. Numerical results are then
presented in Sec.~\ref{sec_res} and we discuss the dependence on the
beam energy and on the principle quantum numbers $n$ and $\kappa$. We
review these results with those existing in the literature in
Sec.~\ref{sec_review}. Finally our conclusions are given in
Sec.~\ref{sec_conclu}.

\section{PWBA or SCA Theory of Bound-Free pair production}
\label{sec_pwba}
The total cross section for bound-free pair production (per
electron-state) in Lorentz gauge is given by
\cite{EichlerM95,MeierHHT98}
\begin{eqnarray}
\sigma^{(bfpp)}_{n_{f},\kappa_{f}} & = & 
8\pi\left(\frac{Z_P \alpha_{em}}{\beta_T}\right)^2
\sum_{\kappa_{i}}\int_{m}^\infty dE_{i}\int_{k_z}^\infty 
\frac{dk}{k^3} \frac{F(k^2-\beta_T^2 k_z^2)}
{\left[1-(\beta_T k_z/ k )^2\right]^2} \nonumber\\
&&\times\sum_{m_fm_i}\left|\left<\psi_f(\vec r)\left|\left(1-
\vec \beta_T \cdot \vec \alpha
\right)e^{i\vec k\cdot\vec r}\right|\psi_i(\vec r)\right>\right|^2,
\label{bf1}
\end{eqnarray}
where $\vec \beta_T=\frac{\vec v}{c}$ is the velocity of the
projectile in the target rest frame,
$\gamma_T=\left(1-\beta_T^2\right)^{-1/2}$ is the Lorentz factor in
the target rest frame, and $F(k^2)$ is the form factor of the charge
distribution of the projectile. For electrons and positrons we can set
this form factor $F(k^2)\equiv 1$.  (For the pair production of muons
and tauons this formfactor will become important.)  We denote the fine
structure constant by $\alpha_{em}$ to distinguish it from the usual
Dirac matrices $\vec\alpha$.  The spatial momentum transfer from
projectile to target is $\hbar \vec k\,$.  The absolute value of the
wave vector $\vec k\,$ is denoted by $k$, and its $z$-component is
related to the energy of the photon as
\begin{equation}
k_z=\frac\omega{v} =\frac{E_f-E_i}{\hbar v}. 
\end{equation}
The total energy of the bound electron is $E_{f}$ and the one of the
electron in the continuum is $E_{i}$.  (Please note that $E_{i}$ is
negative.)  $\psi_i(\vec r\,)$ and $\psi_f(\vec r\,)$ are the
Dirac-Coulomb wave functions describing the initial negative continuum
state and final bound state.  The charge numbers of the projectile and
target are denoted by $Z_P$ and $Z_T$.

The same expression Eq.~(\ref{bf1}) in the Coulomb gauge reads
\begin{eqnarray}
\sigma^{(bfpp)}_{n_{f},\kappa_{f}} & = & 8\pi\left(\frac{Z_P \alpha_{em}}
{\beta_T}\right)^2
\sum_{\kappa_{i}}\int_{m}^\infty dE_{i}\int_{k_z}^\infty \frac{dk}{k^3}
F(k^2-\beta_T^2 k_z^2)
\nonumber\\
&&\times\sum_{m_fm_i}\left|\left<\psi_f(\vec r)\left|\left(1 -
\frac{\vec\beta_{T\perp}\cdot\vec\alpha}{1 - (\beta_T k_z / k)^2}\right)
e^{i\vec k\cdot\vec r}\right|\psi_i(\vec r)\right>\right|^2.
\label{bf2}
\end{eqnarray}
Eq.~(\ref{bf2}) can be obtained from Eq.~(\ref{bf1}) by making use of
the current conservation. Both equations are therefore identical only
when exact eigenfunction of the Dirac Hamiltonian are used, as already
stressed in \cite{EichlerM95,Eichler90}.

The 4-component Dirac spinor in a spherically symmetric field is given 
by
\begin{equation}
\psi_\kappa^m= 
\left(
\begin{array}{c}
g_\kappa(r)\chi_\kappa^m(\hat r) \\
if_\kappa(r)\chi_\kappa^m(\hat r)
\end{array}
\right)
\end{equation}
The angular dependence is expressed by the spin-angular functions
\begin{equation}
\chi_{\kappa}^{m}(\hat r)=\sum_{\tau=\pm\frac12}(-1)^{\ell+m-\frac12}
\sqrt{2j+1}
\left(
\begin{array}{ccc}
\ell & \frac12 & \ \,j \\
m-\tau & \tau & -m
\end{array}
\right)
Y_\ell^{m-\tau}(\hat r)\chi_\tau,
\end{equation}
where $\chi_\tau$ is the Pauli spinor and
\begin{equation}
j=|\kappa|-\frac12,\qquad\ell=j+\frac12\mbox{sgn}(\kappa).
\end{equation}
The radial Dirac equation is
\begin{eqnarray}
\frac {dg}{dr} & = & -\frac{\kappa+1}{r}\,g+\frac{mc}{\hbar}
\left[1+\frac{E-V(r)}{mc^2}\right]f
\label{cd1}\\
\frac {df}{dr} & = &
\frac{\kappa-1}{r}\,f+\frac{mc}{\hbar}
\left[1-\frac{E-V(r)}{mc^2}\right]g.
\label{cd2}
\end{eqnarray}
The radial functions in the Coulomb field 
$V(r)=-\frac{\zeta}{r}$ (with $\zeta=\alpha_{em} Z_T$)
for the bound states are given by
\begin{eqnarray}
\left(
\begin{array}{c}
g_{n,\kappa}(r) \\
f_{n,\kappa}(r)
\end{array}
\right)
& = & N\left[mc^2\pm\sqrt{(mc^2)^2-(\hbar c\beta)^2}\right]^\frac12
(\beta r)^{\gamma-1}e^{-\beta r}
\Biggl[\pm\left(\frac{\zeta mc^2}{\hbar c\beta}-\kappa\right)
\nonumber\\&&\times
\textrm{F}(-n_r,2\gamma+1;2\beta r)-n_r\,\textrm{F}(1-n_r,2\gamma
+1;2\beta r)\Biggr].
\label{rbwf1}
\end{eqnarray}
The corresponding energy eigenvalues are
\begin{equation}
E_{nj}=mc^2\left[1+\frac{\zeta^2}{\left(n-j-\frac12+\gamma
\right)^2}\right]^{-\frac12},
\end{equation}
where $n$ is the principle quantum number defined by
\begin{equation}
n=n_r+|\kappa|.
\end{equation}
The quantity $\beta$ corresponding to the energy eigenvalue is
\begin{equation}
\beta=\frac{mc}{\hbar}\frac{\zeta}
{\left[(n_r+\gamma)^2+\zeta^2\right]^\frac12}.
\end{equation} 
$\gamma$ is given by
\begin{equation}
\gamma=\sqrt{\kappa^2-\zeta^2}.
\end{equation}
The adequate normalization for bound states is 
\begin{equation}
\int d^3r\ \psi^\dagger(\vec r)\,\psi(\vec r)=\int_0^\infty dr\ r^2
\left[g^2(r)+f^2(r)\right]=1
\end{equation}
and the normalization constant is found to be 
\begin{equation}
N=\frac{2^{\gamma}\beta^2}{\Gamma(2\gamma+1)}\left[
\frac{\Gamma(2\gamma+n_r+1)}{2mc^2(n_r!)\,\frac{\zeta mc^2}{\hbar c}
\left(\frac{\zeta mc^2}{\hbar c\beta}-\kappa\right)}\right]^\frac12.
\end{equation}
The continuum Dirac-Coulomb wave functions are
\begin{eqnarray}
\left(
\begin{array}{c}
g_{E,\kappa}(r)\\ f_{E,\kappa}(r)
\end{array}\right) 
& = &\left(\frac{E+mc^2}{E-mc^2}\right)^{\frac 14} 
\frac{k'}{(\pi\hbar c)^\frac12} N_f\,(k'r)^{\gamma-1}
\nonumber\\&&\times
\left(
\begin{array}{c}
\mbox{Re}\\ \mbox{sgn}(E)\sqrt{\frac{E-mc^2}{E+mc^2}}\ \mbox{Im}
\end{array}
\right)
\nonumber\\&&\times
\left[e^{-i(k'r+\varphi)}\,_1
\textrm{F}_1(\gamma+i\eta,2\gamma+1,2ik'r)\right],
\label{rcwf1}
\end{eqnarray}
which are normalized according to
\begin{equation}
\int_0^\infty dr\ r^2\,[g_{E,\kappa}\,g_{E',\kappa}
+f_{E,\kappa}\,f_{E',\kappa}]=\delta(E-E').
\label{nm1}
\end{equation}
In (\ref{rcwf1}) $k'$, $\eta$, $\varphi$ and $N_f$ are given by
\begin{equation}
k'=\frac{\sqrt{E^2-(mc^2)^2}}{\hbar c},\qquad
\eta=\frac{\zeta E}{\hbar ck'},\qquad
e^{2i\varphi}=\frac{-\kappa+i\eta\frac{mc^2}{E}}{\gamma-i\eta},
\end{equation}
\begin{equation}
N_f=\frac{2^{\gamma} 
e^{\frac{\pi \eta} 2}|\Gamma(\gamma+1+i\eta)|}
{\Gamma(2\gamma+1)}.
\end{equation}
Please note that $E>mc^2$ for positive energy continuum wave functions
and $E<-mc^2$ for negative energy states.

Starting from the expression in the Coulomb gauge (\ref{bf2}) the
angular integration can be performed analytically \cite{Meier99} to
get:
\begin{eqnarray} 
\sigma^{(bfpp)}_{n_{f},\kappa_{f}} & = & 32\pi^2\,\left(
\frac{Z_P \alpha_{em}}{\beta_T}\right)^2
\sum_{\kappa_i}\int_m^\infty dE_i\int_{k_z}^\infty \frac{dk}{k^3}
\nonumber\\&&
\times\left\{T_l
+\frac{\beta_T^2}2\,\frac{1}{\left[1-(\beta_T k_z / k)^2\right]^2}
\left(1-\frac{k_z^2}{k^2}\right)T_\perp\right\}.
\label{eq_sigpoint}
\end{eqnarray}
This equation can be rewritten as an integration over $k_\perp$, with
$k^2=k_\perp^2+k_z^2$:
\begin{eqnarray} 
\sigma^{(bfpp)}_{n_{f},\kappa_{f}} & = & 16\pi^2\,\left(
\frac{Z_P \alpha_{em}}{\beta_T}\right)^2
\sum_{\kappa_i}\int_m^\infty dE_i\int_0^\infty \frac{d(k_\perp^2)}
{k_\perp^2+k_z^2}
\nonumber\\&&
\times\left\{\frac{1}{k_\perp^2+k_z^2}T_l
+\frac{\beta_T^2}{2}\,\frac{k_\perp^2}
{\left[k_\perp^2+\left(k_z/\gamma_T\right)^2\right]^2} T_\perp\right\}.
\end{eqnarray}
In this form the increase of the cross section with $\ln\gamma_T$ can
be clearly seen due to the second term (proportional to $T_\perp$) in
the integral. $T_l$ and $T_\perp$ are given by
\begin{eqnarray}
T_l & = & \frac{(2j_f+1)(2j_i+1)}{4\pi}
\sum_\ell (2\ell+1)\frac 12\left[1+(-1)^{\ell_f+\ell+\ell_i}\right]
\nonumber\\
&&\times\left|J^\ell(k)\right|^2
\left(
\begin{array}{ccc}
j_f & \ell & \,j_i \\
\frac 12 & 0 & \!\!\!-\frac 12
\end{array}
\right)^2,
\label{tparallel3}\\
T_\perp & = & \frac{(2j_f+1)(2j_i+1)}{4\pi}\sum_{\ell}(2\ell+1)
\left\{\rule{0pt}{24pt}\right.
\frac 12 \left[1+(-1)^{\ell_f+\ell+\ell_i+1}\right]
\nonumber\\
&&\times\frac 1{\ell(\ell+1)}
\left|\left(\kappa_i+\kappa_f\right)I_\ell^+(k)\right|^2
+\frac 12 \left[1+(-1)^{\ell_f+\ell+\ell_i}\right]\nonumber\\
&&\times\frac 1{(2\ell+1)^2}\left|\rule{0pt}{20pt}\right.\left(
\frac{\ell+1}{\ell}\right)^{\frac 12}
\left[(\kappa_i-\kappa_f)I_{\ell-1}^+(k)-\ell\,I_{\ell-1}^-(k)\right]
\nonumber\\
&&-\left(\frac \ell{\ell+1}\right)^{\frac 12}
\left[(\kappa_i-\kappa_f)I_{\ell+1}^+(k)+(\ell+1)\,I_{\ell+1}^-(k)\right]
\left.\rule{0pt}{20pt}\right|^2\left.\rule{0pt}{24pt}\right\}\nonumber\\
&&\times\left(
\begin{array}{ccc}
j_f & \ell & \,j_i \\
\frac 12 & 0 & \!\!\!-\frac 12
\end{array}
\right)^2,
\label{tsenkrecht3}
\end{eqnarray}
and the radial integrals are
\begin{eqnarray}
J^\ell(k) & = & \int_0^\infty dr\ r^2 j_\ell(kr) 
\left[g_{E_i,\kappa_i}(r)g_{n_{f},\kappa_f}(r)
+f_{E_i,\kappa_i}(r)f_{n_{f},\kappa_f}(r)\right],
\label{jlbf}
\\
I^\pm_\ell(k) & = & \int_0^\infty dr\ r^2 j_\ell(kr)
\left[g_{E_i,\kappa_i}(r)f_{n_{f},\kappa_f}(r)
\pm f_{E_i,\kappa_i}(r)g_{n_{f},\kappa_f}(r)\right]. 
\label{ilbf}
\end{eqnarray}

These rapidly oscillating integrals are very difficult to evaluate.
The integrals $T_l$ and $T_\perp$ are independent of the Lorentz
factor $\gamma$ and can be evaluated before integrating over $k$ in
Eq.~(\ref{eq_sigpoint}). But for high values of $\gamma$ this
integration extends from very low to very high values of $k$, as well
as, $E_i$.  We were able to do the integration for these extreme
limits for RHIC and LHC. For this we have used recursion relations
from \cite{TrautmannBR83}.

\section{Numerical results}
\label{sec_res}

We give our results for symmetrical collisions where the charges of
both ions are equal $Z_{P}=Z_{T}=Z$. The cross sections refer to
capture to {\em one} of the ions {\em only}. Since the cross section 
for capture scales with $Z_{P}^2$, the cross section for asymmetrical 
collisions is obtained by scaling the cross sections with $Z_{P}^2/Z^2$.

In Table \ref{tab_xsec} we give the cross sections for the different
ions and for different bound states. This table was already presented
in \cite{Meier99,MeierHHt99}. We have used the conditions as they are
relevant for Au-Au at RHIC and Pb-Pb at LHC.  In these colliders each
of the ion is assumed to have a Lorentz-factor of $\gamma_c=Z/A
\gamma_p$. For RHIC we take $\gamma_p=250$, for LHC $\gamma_p=7500$,
respectively. For different values of $\gamma_c$ the cross section can
be obtained by using Eq.~(\ref{eq_alnb}).  The cross section are
larger than the ones quoted in the ``ALICE technical report''
\cite{ALICE95}.  The Lorentz factor $\gamma_T$ in the rest frame of
one of the ions, which is the relevant one for the bound-free pair
production, is then given by
\begin{equation}
\gamma_T = 2 \gamma_c^2 -1.
\end{equation}
For RHIC (LHC) this corresponds to $\gamma_T=2\times10^4$
($\gamma_T=1.8\times10^7$).  For large values of $\gamma_c$ it was
found that the cross sections can be parameterized as \cite{BaltzRW93}
\begin{equation}
\sigma=A\,\ln(\gamma_c)+B
\label{eq_alnb}
\end{equation}
We give also the values of the constants $A$ and $B$ in the table.

\begin{figure}
\begin{center}
\ForceHeight{10cm}
\BoxedEPSF{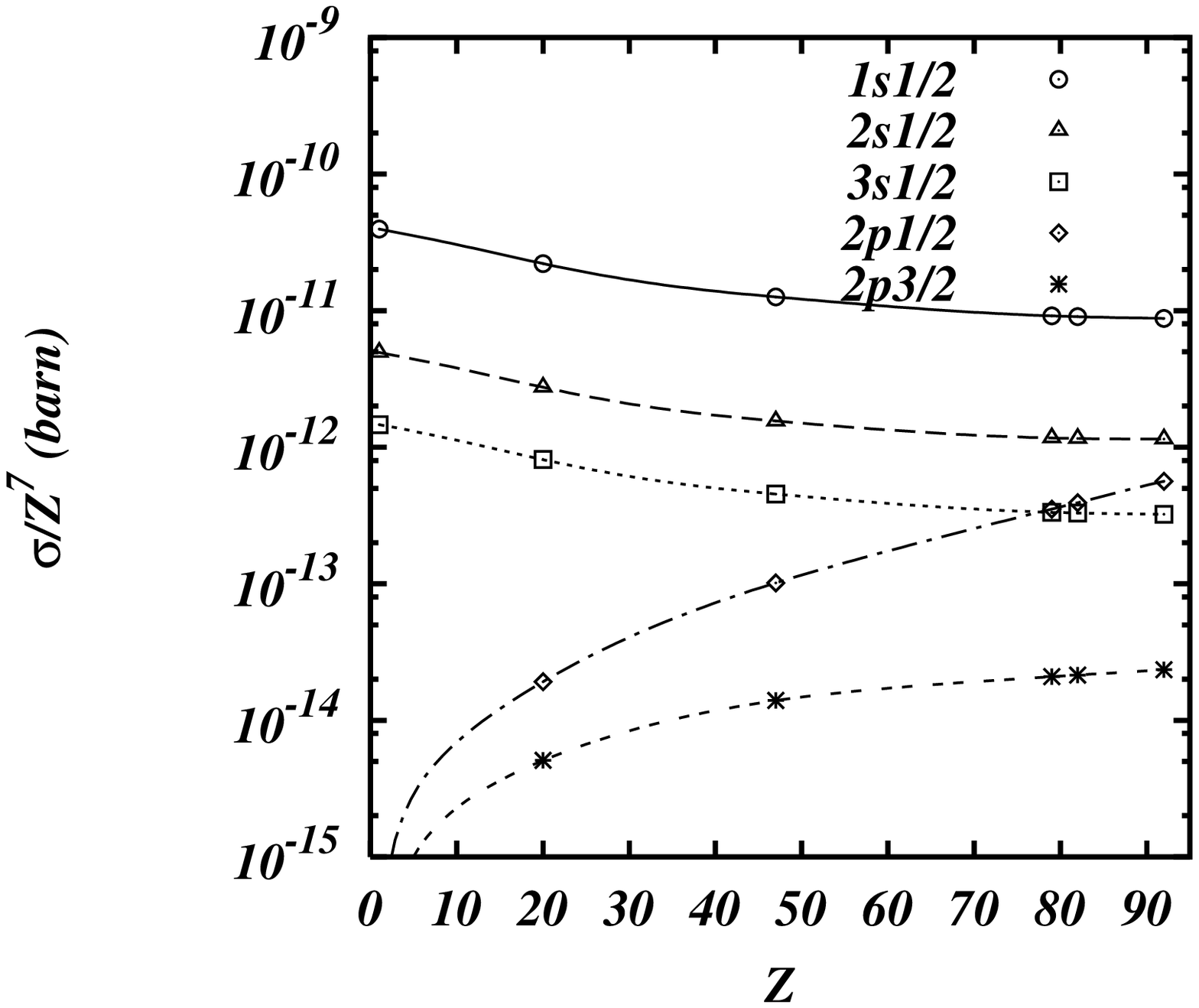}
\end{center}
\caption{The cross section for the capture of the electron into
different bound states is given as a function of the charge of the ion
charge $Z$. A Lorentz factor in the c.m. system of $\gamma_c=3400$,
corresponding to an equivalent Lorentz factor of $\gamma_T=2.3 \times
10^7$ in the rest frame of the ion is used.  For $s$-states there is
an approximate $Z^7$ scaling of the cross section.}
\label{fig_zdep}
\end{figure}

In Figure~\ref{fig_zdep} we show the dependence of the different cross
sections on the ion-charge $Z$. As the cross section for the $s$
states is known to be proportional to $Z^7$ for low values of $Z$
\cite{BertulaniB88}, we divide through this factor.  The most striking
result is that the cross section for the capture into the $2p_{1/2}$
state increases rapidly with $Z$ and becomes even larger than the one
for the $3s$ state for heavy ions. The cross section for the
$2p_{3/2}$ shows a clearly different behavior. We attribute this
difference to the ``small component'' of the Dirac wave function. This
wave function has an $s$-wave character for the $p_{1/2}$ state and a
$d$-wave character for the $p_{3/2}$ state, respectively. With the
increase of $Z$ the ``small component'' gets more important and leads
to the different behavior of the two cross sections.  This is also in
accord with a qualitative discussion of relativistic effects of the
Dirac wave function for bound states as a function of $Z$
\cite{BetheS57}.

For the cross section to the different $s$ states as a function of the
principle quantum number $n$, a scaling law is found in the
nonrelativistic limit,
\begin{equation}
\sigma_{ns}=\frac{\sigma_{1s}}{n^3},
\end{equation}
that also appears, e.g., in the photo-electric effect \cite{BetheS57}.
This scaling law arises because the value of the wave function at the
origin
\begin{equation}
\sigma \sim \left| \psi(0) \right|^2.
\end{equation}
enters in the expression for the cross section.  Our calculations
confirm this $1/n^3$ scaling to a high accuracy once more and for all
values of $Z$. With this scaling law we can sum the contribution from
capture processes into all $s$ states \cite{BertulaniB88}
\begin{equation}
\sum_{n=1}^{\infty} \sigma_{ns} = \zeta(3)\ \sigma_{1s} 
\approx 1.202\  \sigma_{1s}
\end{equation}
where $\zeta$ denotes the Riemann $\zeta$-function.  For low atomic
numbers $Z$ the contributions from $s$-orbits are the most important
ones, whereas the cross section to the $2p_{1/2}$ contributes of the
order of 6.5\% in the heaviest cases.

\section{Comparison with other calculations}
\label{sec_review}
In this section we want to give a comparison of the results that have
been obtained with different approaches. The results for the cross 
section are summarized in Table~\ref{tab_comp}. A large number of
calculations exist including the electromagnetic interaction (with the
projectile) to all orders. Most of them have been done for beam
energies much smaller than those available at the relativistic heavy
ion colliders.  In order to get results for energies at RHIC and
especially at LHC energies extrapolations were used, see below.

A calculation within the PWBA approximation was done by Becker et
al. \cite{AnholtB87,BeckerGS87,Becker87}. Calculations up to values of
$\gamma_T=1000$ have been done and an interpolation formulae of the
form
\begin{equation}
\sigma \approx Z_P^2 Z_T^5 a \ln\left(\gamma_T/\gamma_0\right)
\end{equation}
(using our notation) and tabulated values for $a$ and $\gamma_0$ can
be found in \cite{AnholtB87}.  The parameters $a$ and $\gamma_0$
depend slightly on $Z_T$.  Using the value of $a$ for $Z=80$ for both
Au and Pb, we get values of 93~barn (Au-Au at RHIC) and 226~barn
(Pb-Pb at LHC), respectively.

In \cite{BertulaniB88} a calculation is done making use of the
Sommerfeld-Maue wave functions as an approximation to the exact
Coulomb-Dirac continuum wave functions and using also an approximation
for the bound state wave function, which is valid in the limit
$Z\alpha\ll1$. They found their result in SCA to be equal to the one
derived within the equivalent photon approximation.  Using their
formula we obtain results which are substantially lower (by a factor
of two) than our results here. As their approximation is strictly
valid only in the limit $Z\alpha\ll1$, such a discrepancy is perhaps
not surprising.

Baltz et al. treated the problem in a series of papers
\cite{BaltzRW91,BaltzRW93,BaltzRW94,BaltzRW96}. For large impact
parameter they also make use of a perturbative treatment. In addition
they simplify the interaction potential by taking only lowest order
terms in $1/\gamma_T$ and also expanding in $\rho/b$, where $\rho$ is
the electron coordinate and $b$ the impact parameter. They proposed
the parameterization of the cross section for large values of
$\gamma_T$ of the form
\begin{equation}
\sigma = A \ln(\gamma_T) + B,
\label{eq_baltz}
\end{equation}
with the interpretation of $A$ to be given by the perturbative part
only and the influence of higher order terms at small impact parameter
to be present in $B$ only (Of course this parameterization is
identical to the one of Becker et al.  above).  Tabulated values for
$A$ and $B$ are given in \cite{BaltzRW93}. From them (and choosing for
$b_{min}=1 \lambda_c$) we get cross section of 83 barn (Au-Au at
RHIC), 161~barn (Au-Au at LHC), and 436~barn (U-U at LHC). They also
give the cross section to be 89 barn (Au-Au at RHIC) in the text where
contributions from nonperturbative processes at small impact parameter
have been included. No results for lead-lead collisions are given, but
assuming a $Z^7$ scaling we get from the extrapolation of the Au
results a value of 206 barn (Pb-Pb at LHC), from the extrapolation of
the U results 195 barn (Pb-Pb at LHC) using the $\gamma_T$ dependence
of Eq.~(\ref{eq_baltz}).

In \cite{RhoadesBrownBS89} a calculation is done using the cross
section of the free pair production cross section and folding it with
the momentum distribution of the bound state. The electron therefore
is described by a plane wave. For Au-Au collisions at RHIC a cross
section of 72~barn was found by them, but see also \cite{Baur91b}.

Two calculations making use of the equivalent photon approximations
have been performed also. In both exact solutions of the Dirac
equations are used.  In the equivalent photon method a cutoff
parameter has to be introduced, which leads to some uncertainty in the
results.  In \cite{AsteHT94} capture to the $1s$-state was
calculated. The results were also compared with the cross section for
bound-free pair production induced by real photons at low photon
energies by \cite{JohnsonBC64}.  For the heavy ion case values of 90
barn (Au-Au at RHIC) and 222~barn (Pb-Pb at LHC, $\gamma_c=3400$) are
given, respectively.  Also tabulated values of $A$ and $B$ are
presented. Their results are in good agreement with the present
values.  The cutoff parameter was taken to be the Compton wavelength
of the electron.  In \cite{AggerS97} similar calculations are
performed including also capture cross section calculations to the
$L$-shell. Agreement with the results of \cite{AsteHT94} was found for
the $K$-shell. A critical discussion of the use of the
cutoff-parameter present in the equivalent photon spectrum was done.
The exact value of this parameter is not given by the
Weizs\"acker-Williams theory and therefore introduces some
uncertainties in the results. For Au-U collisions at RHIC with the
electron being captured by the U-atom they find a value of 165~barn
for $b_{min}=2\lambda_c$ and 182~barn for $b_{min}=\lambda_c$ with
$\lambda_c$ the Compton wavelength of the electron. Comparing these
results with the 195~barn we get, an cutoff parameter of
$b_{min}=\lambda_c$ seems to be favored.  Again assuming a scaling of
the result with $Z_T^5$ we get for Au-Au collisions at RHIC 77~barn
($b_{min}=2\lambda_c$), 85~barn ($b_{min}=\lambda_c$),
respectively. Unfortunately no results for LHC energies are given.

Experiments were first done at Bevalac at 1~GeV/A
\cite{Belkacem93,Belkacem94,Belkacem97}. A cross section of
2.19(0.25)~barn for Au-U collisions and capture on the U projectile
was found. Experiments were also done at the AGS at 10.8~GeV/A
corresponding to $\gamma_T=12.6$, $\gamma_c=2.6$
\cite{Claytor97,Belkacem98}. A cross section of 8.8~barn was found for
Au-Au collisions (Our calculation gives 11.86~barn\footnote{Please 
note that while we compare cross section for the
$1s$-state only in the comparison with other calculations, comparing
with experimental results we use the total cross section into all
bound states.} The experimental results were found to be in agreement
with the theoretical results of \cite{BeckerGS87,RhoadesBrownBS89}).

A measurement has recently been done at the CERN SPS with a 158 GeV/A
Pb beam ($\gamma_T\approx168$, corresponding to $\gamma_c=9.2$)
\cite{Grafstroem99}, see also \cite{Krause98}. The capture to the Pb
ion was studied for several different targets. For a gold target a
value of 44.3~barn is given (our result is 45~barn). Assuming a
scaling of the cross section as given above Eq.~(\ref{eq_alnb}) and
assuming $B=-24$~barn from \cite{BaltzRW94}, they give extrapolated
values of 94~barn (Au at RHIC) and 204~barn (Pb at LHC). These values
again agree quite well with our results.

\section{Conclusions and Outlook}
\label{sec_conclu}
Apart from neglecting higher order effects in the projectile charge,
we have given a full calculation of the bound-free capture
cross-section.  Since such higher order effects were shown to be small
\cite{Baltz97}, our predictions should be very reliable.  Overall our
results are in agreement with a number of other calculations, using
different approaches.  These calculations are also very important for
the questions of luminosity loss and localized beam pipe heating
\cite{Klein00}, especially at the LHC.  

Muon and $\tau$-lepton bound-free pair-production can be calculated
similarly.  Since the bound-free pair-production cross section scales
with $1/m_{lepton}^2$ \cite{BertulaniB88} these cross-sections will be
much smaller than the ones for $e^+$-$e^-$ pair-production. Especially
for heavy nuclei, it will be necessary to use finite size lepton wave
functions. In addition to the $1/m_{lepton}^2$ scaling there will be
also a severe reduction of the cross section due to the formfactor
effects. Maybe such exotic atoms could be interesting for physics
\cite{Weiss85}? We mention that the lifetime of the free tau-lepton is
given by $ct=87.2\mu$. For a Lorentz factor of about 100 this is still
a very short distance of about 0.1mm for the decay length of such an
atom. On the other hand a muonic atom lives long enough and could be
extracted from the beam (similar to the antihydrogen experiments
\cite{BaurO96,Blanford98}).

In a fixed target experiment, the effect of screening of the atomic
electrons can be incorporated as well by using a screened Coulomb
potential for both the wave function, as well as, the interaction with
the projectile ($F\not\equiv1$ in Eq.~(\ref{bf1})).  This could be
relevant for the CERN fixed target experiments
\cite{Grafstroem99,Krause98}. This effect has been studied in
\cite{VoitkivGS00} and was found to be on the percent level.

RHIC is running right now and we expect that the present numbers will
soon be tested.  After the preparation of this manuscript, an article
by Bertulani and Dolci appeared \cite{BertulaniD00}. This is a
continuation of \cite{BertulaniB98} to the heavy ion case. In it they
give an explanation why the approximation used in \cite{BertulaniB88}
does not work well for large values of $Z$.

\acknowledgements
We would like to thank Daniel Brandt and Spencer Klein for very interesting
and useful comments and discussions, which helped us also in order to make 
this work more clear.

\mediumtext
\begin{table}[h]
\begin{center}
\begin{tabular}{cllll} 
bound state &$\sigma($RHIC$)[b]$&$\sigma($LHC$)[b]$ 
& $A$[b] & $B$[b] \\
\hline
{$^{1}$H$-^{1}$H} & $\gamma_c=250$ & $\gamma_c=7500$ & & \\
$1s$      &$2.62\cdot10^{-11}$ & $4.25\cdot10^{-11}$ 
& $5.36\cdot10^{-12}$ & $-3.40\cdot10^{-12}$ \\
$2s$      &$3.28\cdot10^{-12}$ & $5.31\cdot10^{-12}$ 
& $6.70\cdot10^{-13}$ & $-4.23\cdot10^{-13}$ \\
$2p(1/2)$ &$3.75\cdot10^{-17}$ & $6.10\cdot10^{-17}$ 
& $7.73\cdot10^{-18}$ & $-5.20\cdot10^{-18}$ \\
$2p(3/2)$ &$1.47\cdot10^{-17}$ & $2.41\cdot10^{-17}$ 
& $3.10\cdot10^{-18}$ & $-2.42\cdot10^{-18}$ \\
$3s$      &$9.70\cdot10^{-13}$ & $1.57\cdot10^{-12}$ 
& $1.98\cdot10^{-13}$ & $-1.26\cdot10^{-13}$ \\
\hline
{$^{20}$Ca$-^{20}$Ca} & $\gamma_c=125$ & $\gamma_c=3750 $ & & \\
$1s$      &$1.61\cdot10^{-2}$ & $2.92\cdot10^{-2}$  
& $3.84\cdot10^{-3}$ & $-2.48\cdot10^{-3}$ \\
$2s$      &$2.00\cdot10^{-3}$ & $3.62\cdot10^{-3}$  
& $4.78\cdot10^{-4}$ & $-3.07\cdot10^{-4}$ \\
$2p(1/2)$ &$1.39\cdot10^{-5}$ & $2.52\cdot10^{-5}$  
& $3.35\cdot10^{-6}$ & $-2.33\cdot10^{-6}$ \\
$2p(3/2)$ &$3.63\cdot10^{-6}$ & $6.70\cdot10^{-6}$  
& $9.02\cdot10^{-7}$ & $-7.27\cdot10^{-7}$ \\
$3s$      &$5.90\cdot10^{-4}$ & $1.07\cdot10^{-3}$  
& $1.41\cdot10^{-4}$ & $-9.10\cdot10^{-5}$ \\
\hline
{$^{47}$Ag$-^{47}$Ag} & $\gamma_c=109$ & $\gamma_c=3264$ & & \\
$1s$      &$3.51$             & $6.46$              
& $8.68\cdot10^{-1}$ & $-5.63\cdot10^{-1}$ \\
$2s$      &$4.33\cdot10^{-1}$ & $7.98\cdot10^{-1}$  
& $1.07\cdot10^{-1}$ & $-6.94\cdot10^{-2}$ \\
$2p(1/2)$ &$2.81\cdot10^{-2}$ & $5.21\cdot10^{-2}$  
& $7.05\cdot10^{-3}$ & $-5.02\cdot10^{-3}$ \\
$2p(3/2)$ &$3.80\cdot10^{-3}$ & $7.16\cdot10^{-3}$  
& $9.87\cdot10^{-4}$ & $-8.31\cdot10^{-4}$ \\
$3s$      &$1.26\cdot10^{-1}$ & $2.34\cdot10^{-1}$  
& $3.13\cdot10^{-2}$ & $-2.02\cdot10^{-2}$ \\
\hline
{$^{79}$Au$-^{79}$Au} & $\gamma_c=100$ & $\gamma_c=3008$ & & \\
$1s$      &$94.9$             & $176$               
& $23.8$             & $-14.7$            \\
$2s$      &$12.1$             & $22.4$              
& $3.04$             & $-1.87$            \\
$2p(1/2)$ &$3.62$             & $6.77$              
& $9.27\cdot10^{-1}$ & $-6.56\cdot10^{-1}$ \\
$2p(3/2)$ &$2.10\cdot10^{-1}$ & $4.01\cdot10^{-1}$  
& $5.62\cdot10^{-2}$ & $-4.93\cdot10^{-2}$ \\
$3s$      &$3.46$             & $6.40$              
& $8.67\cdot10^{-1}$ & $-5.34\cdot10^{-1}$ \\
\hline
{$^{82}$Pb$-^{82}$Pb} & $\gamma_c=99$ & $\gamma_c=2957$ & & \\
$1s$      &$121$              & $225$               
& $30.4$  & $-18.7$            \\
$2s$      &$15.5$             & $28.8$              
& $3.91$  & $-2.39$            \\
$2p(1/2)$ &$5.21$             & $9.76$              
& $1.34$  & $-9.46\cdot10^{-1}$ \\
$2p(3/2)$ &$2.78\cdot10^{-1}$ & $5.33\cdot10^{-1}$  
& $7.50\cdot10^{-2}$ & $-6.61\cdot10^{-2}$ \\
$3s$      &$4.42$             & $8.20$              
& $1.11$ & $-6.79\cdot10^{-1}$ \\
\hline
{$^{92}$U$-^{92}$U} & $\gamma_c=97$ & $\gamma_c=2900$ & & \\
$1s$      &$263$              & $488$               
& $66.0$  & $-39.0$ \\
$2s$      &$34.4$             & $63.7$              
& $8.63$  & $-5.10$ \\
$2p(1/2)$ &$16.7$             & $31.3$              
& $4.30$  & $-3.00$ \\
$2p(3/2)$ &$6.77\cdot10^{-1}$ & $1.30$              
& $1.83\cdot10^{-1}$ & $-1.63\cdot10^{-1}$ \\
$3s$      &$9.67$             & $17.9$              
& $2.43$  & $-1.44$ \\
\end{tabular}
\caption{Cross section for the bound-free pair production of {\em one } ion
{\em only} for different bound states are given for RHIC and LHC conditions 
for different ion-ion collisions. Also given are the parameters $A$ and
$B$ to be used in Eq.~(\protect\ref{eq_alnb}) for the dependence on
the Lorentz factor $\gamma_c$.}
\label{tab_xsec}
\end{center}
\end{table}

\narrowtext
\begin{table}
\begin{tabular}{ccc}
$\sigma$(Au-Au,RHIC,$\gamma_c=100$) [barn] 
& $\sigma$(Pb-Pb,LHC,$\gamma_c=2957$) [barn] &Ref. \\
94.9 & 225 & This work \\
\hline
93   & 226 & \protect\cite{AnholtB87,Becker87}\\
\hline
37   & 86  & \protect\cite{BertulaniB88}\\
\hline
89   & 206 (from U)  & \protect\cite{BaltzRW91,BaltzRW93,BaltzRW94}\\
     & 195 (from Au) & \\
\hline
72   & ---           & \protect\cite{RhoadesBrownBS89}\\
90   & 222 ($\gamma_c=3400$) & \protect\cite{AsteHT94}\\
77 ($b_{min}=2\lambda_c$)  & ---  & \protect\cite{AggerS97}\\
85 ($b_{min}=\lambda_c$)   &      &\\
94   & 204 barn                   & \protect\cite{Grafstroem99}
\end{tabular}
\caption{Cross section for bound-free pair production of {\em one} ion
{\em only} into the $1s$-state calculated by different groups are
compared. The last entry uses experimental results of bound-free pair
production into any bound state at CERN SPS with 158~GeV/A
energies. Given are the results for Au-Au collisions at RHIC and Pb-Pb
collisions at LHC. For details of the different calculations see the
text.}
\label{tab_comp}
\end{table}
\end{document}